\documentstyle[preprint,aps]{revtex}

\begin{document}

\draft

\preprint{November,2000}

\title{Znajek-Damour Horizon Boundary Conditions with \\
       Born-Infeld Electrodynamics}

\author{Hongsu Kim\footnote{e-mail : hongsu@hepth.hanyang.ac.kr},
Hyun Kyu Lee\footnote{e-mail : hklee@hepth.hanyang.ac.kr} and 
Chul H. Lee\footnote{e-mail : chlee@hepth.hanyang.ac.kr}}

\address{Department of Physics \\
Hanyang University, Seoul, 133-791, KOREA}


\maketitle

\begin{abstract}
In this work, the interaction of electromagnetic fields with a rotating (Kerr) black hole 
is explored in the context of Born-Infeld (BI) theory of electromagnetism instead of 
standard Maxwell theory and particularly BI theory versions of the four horizon boundary 
conditions of Znajek and Damour are derived. Naturally, an issue to be addressed is then 
whether they would change from the ones given in the Maxwell theory context and if they 
would, how. Interestingly enough, as long as one employs the same local null tetrad frame 
as the one adopted in the works by Damour and by Znajek to read out physical values of 
electromagnetic fields and fictitious surface charge and currents on the horizon, 
it turns out that one ends up with exactly the same four horizon boundary conditions 
despite the shift of the electrodynamics theory from a linear Maxwell one to a highly 
non-linear BI one. Close inspection reveals that this curious and unexpected result 
can be attributed to the fact that 
the concrete structure of BI equations happens to be such that it is indistinguishable 
{\it at the horizon} to a local observer, say, in Damour's local tetrad frame from that 
of standard Maxwell theory.

\end{abstract}


\narrowtext

\begin{center}
{\rm \bf I. Introduction}
\end{center}

The idea of using rotating black holes as energy sources has a long history. To our
knowledge, Salpeter [1] and Zel'dovich [1] were the first to point out that gigantic
black holes might serve as power engines for quasars or radio galaxies. Realistic
theoretical models to realize this type of energy extraction from rotating black holes
also appeared afterwards and they are due to Penrose [2], Press and Teukolsky [2],
Ruffini and Wilson [3], Damour [3], and Blandford and Znajek [4]. 
Among these models, that of Blandford and Znajek is 
particularly interesting in its formulation and looks quite plausible in its operational
mechanism. At first, puzzling over the possible explanation for the observed twin
jets pointing oppositely out of a black hole-accretion disk system, Blandford and
Znajek conceived of a particular process in which the power going into the jets comes
from the hole's enormous rotational energy. Schematically, their mechanism works as 
follows ; suppose that the rotating hole is threaded by magnetic field lines. As the
hole spins, it drags the field lines around, causing them to fling surrounding plasma
upward and downward to form two jets. Then the jets shoot out along the hole's spin
axis and their direction is firmely fixed to the hole's axis of rotation. The magnetic
field lines, of course, come from the accretion disk around the hole. Namely, it is
the magnetic fields that extract the rotational energy of a black hole and then act
to power the jets. According to their careful analysis, on the other hand, as the 
energy is extracted, electric currents flow into the horizon near the hole's poles 
(in the form of positively-charged particles falling inward), and currents flow out
of the horizon near the equator (in the form of negatively-charged particles falling 
inward). It was as though the hole were a voltage generator of an electric circuit
driving current out of the horizon's equator, then up magnetic field lines to a
large distance, then through ``plasma load'' to other field lines near the hole's
spin axis, then down those field lines and into the horizon. Namely, the magnetic field 
were the wires of the electric circuits, the plasma was the load that exerts power
from the circuit. And the two pictures, one schematic and the other analytic, are
just two different ways of describing the same phenomenon. This electric circuit
description was totally unexpected and thus curious enough although it was resulted from a
careful general relativistic treatment of the problem. Right after the post of this
new mechanism, Znajek [5] and, independently, Damour [6] succeeded in translating the careful
general relativistic formulation into a surprisingly simple non-relativistic, flat
spacetime electrodynamics language, the celebrated four horizon boundary conditions. 
And the assumption of central importance in this new picture is to endow the horizon
with some fictitious surface charge and current as those previously imagined by
Hanni and Ruffini [7]. It is really amusing that one now has an option to view the
situation in terms of flat spacetime electrodynamics alone at least for rough
understanding. \\
Speaking of the theory that governs the electromagnetism, however, it is interesting to
note that historically, there has been another classical theory that can be thought of as
a larger class of theory involving the standard Maxwell theory just as its limiting case.
It is the theory proposed in the 1930's by Born and Infeld [9].
In spite of its long history, the Born-Infeld (BI) theory of electrodynamics has remained almost
unnoticed and hence nearly uncovered in full detail.
This theory may be thought of as a highly nonlinear generalization of or
a non-trivial alternative to the standard Maxwell theory of electromagnetism. It is known that
Born and Infeld had been led, when they first constructed this theory, by the considerations
such as finiteness of the energy in electrodynamics, natural recovery of the usual Maxwell
theory as a linear approximation and the hope to find soliton-like solutions representing
point-like charged particles. In the present work we would like to explore the interaction
of electromagnetic fields with a rotating (Kerr) black hole but in the context
of BI theory of electromagnetism instead of Maxwell theory. And our particular concern
is to derive BI theory versions of the four horizon boundary conditions to see how they
would change from the ones derived originally by Znajek and by Damour 
in the context of Maxwell theory. 
Now the motivation for shifting the theory of electromagnetism from that of standard
Maxwell to that of BI to study the physics of interaction between ``test''
electromagnetic field and ``background'' rotating black hole geometry can be
stated as follows.
The BI theory, although appeared as a ``classical'' theory long before the advent of
Quantum Electrodynamics (QED) theory, may be viewed as some kind of an effective
low-energy theory of QED in that its highly non-linear structure plays the role of
eliminating the short-distance divergences. Normally, the strong magnetic field,
believed to be anchored in the central black holes of typical gamma-ray bursters,
is regarded as being originated, say, from that of neutron stars that has collapsed to form
the black hole.  A number of various observations
indicate that in young neutron stars, the surface magnetic field strengths are of order
$10^{11} \sim 10^{13} (Gauss)$ and in some extreme cases such as magnetars, magnetic
field strengths are estimated to be as large as $\geq 5\times 10^{14} (Gauss)$ [8]. Then the
magnetic field of this ultra strength, in turn, stimulates our curiosity and leads us
to ask questions such as ; what would happen if we choose to employ the BI theory
that, as stated, can be thought of as an effective theory of QED, instead of linear
Maxwell theory, to stdudy the physics in the vicinity of rotating hole's horizon ?
And in doing so, we anticipate that perhaps the highly non-linear nature of the BI
theory may serve to uncover some hidden interplay between the strong electromagnetic
field and ultra strong gravity near the hole's horizon. Since our main concern is
the derivation of the four horizon boundary conditions in BI theory, we now recall
some of the basic ingredients of these boundary conditions obtained in the conventional
Maxwell theory. \\
The four ``horizon boundary conditions'' first derived in the works of Znajek [5] 
and of Damour [6] and reformulated later in the literature can be briefly described
as follows. They may be called radiative ingoing boundary condition, Ohm's law,
Gauss' law and Ampere's law, respectively. And in order to represent each boundary
condition properly, we need to introduce in advance some quantities that will be derived 
carefully in the text shortly. They are electric and magnetic fields at the horizon
($\vec{E}_{H}$, $\vec{B}_{H}$) as  seen by a local observer in a null tetrad frame which has 
been made to be well-behaved at the horizon by the amount of  boost that becomes 
suitably infinite at the horizon and the {\it fictitious} charge and current densities
($\sigma$, $\vec{\kappa}$) that have been assigned at the horizon in such a way that 
the sum of real current 4-vector outside the horizon and this fictitious current 4-vector 
on the horizon together is conserved. Firstly, then, the radiative ingoing boundary condition
first derived by Znajek [5] takes the form $\vec{B}_{H} = \vec{E}_{H} \times \hat{n}$ with
$\hat{n}$ being the outer unit normal to the horizon. Evidently, it states that the electric 
and magnetic fields tangential to the horizon are equal in magnitude and perpendicular 
in direction and hence their Poynting energy flux is {\it into} the hole.
Secondly, the Ohm's law reads $\vec{E}_{H} = 4\pi \vec{\kappa}$. It has been derived rigorously
in the work by Damour [6] and pointed out in the work by Znajek [5]. Clearly, this relation 
takes on the form of a non-relativistic Ohm's law for a conductor and hence implies that 
if we endow the horizon with some charge and current densities (which are to be determined 
by the surrounding external electromagnetic field $F_{\mu\nu}$ as we shall see in the text), 
then the horizon behaves as if it is a conductor with finite surface resistivity of
$\rho = 4\pi \simeq 377 ({\rm ohms})$. Actually these two relations are the ones that
have been explicitly derived in the works by Znajek and by Damour and play the central role 
in justifying that the introduction of fictitious charge and current densities on the horizon
indeed provides a self-consistent picture. That is, one might wonder what would happen to 
the Joule heat generated when those surface currents work against the surface resistance 
and how it would be related to the electromagnetic energy going down the hole through the 
horizon. In their works, Znajek and Damour provided a simple and natural answer to this question. 
Namely, they showed in an elegant manner that the total electromagnetic energy flux 
(i.e., the Poynting flux) into the rotating Kerr hole through the horizon is indeed precisely 
the same as the amount of Joule heat (Ohmic dissipation) produced by the surface currents 
when they work against the surface resistivity of $4\pi $. As a result, one may think of the
rotating hole as a conducting sphere that absorbs the incident electromagnetic energy
flux as a form of Joule heat that the surface current (driven by the electromagnetic fields)
generates when it interferes with the surface resistivity. This is indeed an interesting and
quite convincing alternative picture of viewing the interaction of external electromagnetic fields 
with a rotating black hole.  Damour [6] also remarked that this result provides a clear confirmation
of Carter's assertion [10] that a black hole is analogous to an ordinary object having finite
viscosity and electrical conductivity. Thirdly, if one follows the formulation of Damour but
in a slightly different way in taking the local tetrad frame and projecting the Maxwell
field tensor and the surface current 4-vector onto that chosen tetrad frame, one also gets the
relation $E_{\hat{r}} = 4\pi \sigma$
which may be identified with the surface version of Gauss' law. It says that the
fictitious surface charge density we assumed on the horizon plays the role of terminating
the normal components of all electric fields that pierce the horizon.
Lastly, if we combine the radiative ingoing boundary condition at the horizon that we
obtained earlier, $\vec{B}_{H} = \vec{E}_{H} \times \hat{n}$ with the Ohm's law 
$\vec{E}_{H} = 4\pi \vec{\kappa}$, we end up with the fourth relation
$\vec{B}_{H} = 4\pi (\vec{\kappa} \times \hat{n})$
which can be viewed as the surface version of Ampere's law. Again, consistently with our
motivation for introducing fictitious current density on the horizon, this relation
indicates that the current density we assumed plays the role of terminating any tangential
components of all magnetic fields penetrating the horizon. And actually these four
horizon boundary conditions later on provided a strong motivation for the proposal of
so-called ``membrane paradigm [11]'' of black holes by Thorne and his collaborators.
As we already mentioned, in the present work we would like to particularly 
derive BI theory versions of these four horizon boundary conditions to see if they
would change from the ones given above and if they would, how. Interestingly enough,
as far as we employ the same local null tetrad frame as the one adopted in the
works by Damour and by Znajek, it turns out that we end up with exactly the same four
horizon boundary conditions despite the shift of the electrodynamics theory
from a linear Maxwell one to a highly non-linear BI one. As we shall see shortly in
the text, this curious and unexpected result can be attributed to the fact that the nature 
of the BI theory or more precisely, the concrete structure of BI equations happens to be 
such that it is indistinguishable {\it at the horizon} to a local observer, say, in Damour's 
local tetrad frame from that of standard Maxwell theory. We find this point indeed quite amusing 
on theoretical side.

\begin{center}
{\rm \bf II. Choice of coordinate system and tetrad frame}
\end{center}

The relevant choice of coordinate system and the proper choice and treatment of the associated tetrad 
frame for the background Kerr black hole spacetime is of primary importance to discuss electrodynamics
on this geometry in terms of {\it physical} field values (here, the meaning of ``physical'' values will
be unambiguously defined shortly). Thus in this section, we shall carefully choose the coordinate
system and perform a proper treatment of the associated tetrad frame to derive the four horizon
boundary conditions based on these choices later on. Generally speaking,
in order to represent the background Kerr geometry, we need to choose a coordinate 
system in which the metric is to be given and in order to obtain physical components
of a tensor (such as the electric and magnetic field values), we need to select a
tetrad frame (in a given coordinate system) to which the tensor components are to be
projected. It has been known for some time that there are three important coordinate
systems for Kerr spacetime ; ingoing/outgoing Kerr coordinates, Kerr-Schild coordinates,
and Boyer-Lindquist coordinates. First, the Kerr-Schild coordinates are quasi-Cartesian
coordinates and the well-known ring structure of curvature singularity can only be
uncovered in this coordinate system. Next, the Boyer-Lindquist coordinates can be
viewed as the generalization of Schwarzschild coordinates to the stationary,
axisymmetric case. Lastly, the Kerr coordinates can be thought of as the axisymmetric
generalization of Eddington-Finkelstein null coordinates and hence are free of coordinate
singularities. In particular, the ingoing (advanced) null coordinates represent a
reference frame of ``freely-falling'' photons. As such, these ingoing Kerr coordinates
are well-behaved on the event horizon and thus meet our purpose to explore the 
electrodynamics near the horizon. Turning to the choice of tetrad frame, there are
largely two types of tetrad frames ; orthonormal tetrad and null tetrad. As is well-known,
the orthonormal tetrad is a set of four mutually orthogonal unit vectors at each
point in a given spacetime which give the directions of the four axes of locally-
Minkowskian coordinate system
\begin{eqnarray}
ds^2 &=& g_{\mu\nu}dx^{\mu}dx^{\nu} = \eta_{AB}e^{A}e^{B} \nonumber \\
&=& -(e^0)^2 + (e^1)^2 + (e^2)^2 + (e^3)^2 
\end{eqnarray}
where $e^{A}=e^{A}_{\mu}dx^{\mu}$. Namely, every physical observer with 4-velocity
$u^{\mu}$ has associated with him an orthonormal frame in which the basis vectors are
the (reciprocal of) orthonormal tetrad $e_{A}=\{e_{0}=u, ~e_{1}, ~e_{2}, ~e_{3}\}$.
And corresponding to this is a null tetrad $Z_{A}=\{l, ~n, ~m, ~\bar{m}\}$
defined by
\begin{eqnarray}
e_{0} &=& {1\over \sqrt{2}}(l + n), ~~~e_{1} = {1\over \sqrt{2}}(l - n), \\
e_{2} &=& {1\over \sqrt{2}}(m + \bar{m}), ~~~e_{3} = {1\over \sqrt{2}i}(m - \bar{m})
\nonumber
\end{eqnarray}
satisfying the orthogonality relation
\begin{eqnarray}
-l^{\mu}n_{\mu} = 1 = m^{\mu}\bar{m}_{\mu}
\end{eqnarray}
with all other contractions being zero and
\begin{eqnarray}
g^{\mu\nu} = - l^{\mu}n^{\nu} - n^{\mu}l^{\nu} + m^{\mu}\bar{m}^{\nu}
+ \bar{m}^{\mu}m^{\nu}.
\end{eqnarray}
Conversely, given a non-singular null tetrad, there is a corresponding physical
observer. The tetrad vectors then can be used to obtain, from tensors in
arbitrary coordinate system, their physical (i.e., finite and non-zero) components
measured by an observer in this locally-flat tetrad frame. And the rules for
calculating the physical components of a tensor, say, $T_{\mu\nu}$ in the orthonormal
frame and in the null frame are given respectively by
\begin{eqnarray}
T_{AB} = T_{\mu\nu}(e_{A}^{\mu}e_{B}^{\nu}),
~~~T_{lm} = T_{\mu\nu}(l^{\mu}m^{\nu}), ~~~{\rm etc.}
\end{eqnarray}
where $e_{A}^{\mu}$ is the inverse of the tetrad vectors $e^{A}_{\mu}$ in that
$e_{A}^{\mu}e^{A}_{\nu} = \delta^{\mu}_{\nu}$ and $e_{A}^{\mu}e^{B}_{\mu} = \delta_{A}^{B}$.

{\rm \bf 1. Hawking-Hartle (or Teukolsky) tetrad}

As we stated earlier in the introduction, we would like to derive Znajek-Damour-type 
boundary conditions at the horizon of Kerr black hole in the context of BI theory of
electromagnetism. Generally speaking, all that is required of the ``correct'' boundary
conditions for electric and magnetic fields at the horizon can be stated as follows.
The physical field's components in the neighborhood of an event horizon should have
``nonspecial'' values. Or put another way, a physically well-behaved observer at the 
horizon should see the fields as having finite and non-zero values. And this can be 
achieved only when one works in the coordinates having non-singular behavior at the
horizon with the choice of a null tetrad frame. Such a choice of well-behaved null
tetrad frame has been provided long ago by Hawking and Hartle [12] and also by
Teukolsky [13]. Thus in order to briefly review the derivation of their tetrad, we
start with Kinnersley's null tetrad [14] given originally in Boyer-Lindquist 
coordinates $x^{\mu}=(t,r,\theta,\tilde{\phi})$ 
\begin{eqnarray}
l^{\mu} &=& \left({(r^2+a^2)\over \Delta}, ~~1, ~~0, ~~{a\over \Delta}\right), \nonumber \\
n^{\mu} &=& \left({(r^2+a^2)\over 2\Sigma}, ~~{-\Delta \over 2\Sigma}, ~~0, 
~~{a\over 2\Sigma}\right), \\
m^{\mu} &=& {1\over \sqrt{2}(r+ia \cos \theta)}\left(ia\sin \theta, ~~0, ~~1, 
~~{i\over \sin \theta}\right). \nonumber
\end{eqnarray}
where $\Sigma = r^2+a^2\cos^2\theta $ and $\Delta = r^2+a^2-2Mr $ with $M$ and $a$
being the ADM mass and the angular momentum per unit mass of the hole respectively.
As is well-known, this tetrad is not well-behaved on the horizon where $\Delta = 0$
since it is given in Boyer-Lindquist coordinates (which themselves are singular on
the horizon) and hence cannot be of any practical use. Thus we transform them to
the ingoing Kerr coordinates $x^{'\mu}=(v,r,\theta,\phi)$ via the coordinate 
transformation law
\begin{eqnarray}
dv = dt + {(r^2+a^2)\over \Delta}dr,
~~~d\phi = d\tilde{\phi} + {a\over \Delta}dr
\end{eqnarray}
to obtain, after the standard procedure,
\begin{eqnarray}
Z^{'\mu}_{A} = \left({\partial x^{'\mu}\over \partial x^{\nu}}\right)Z^{\nu}_{A}
~~~{\rm where} ~~~Z^{\mu}_{A}=(l^{\mu}, n^{\mu}, m^{\mu}, \bar{m}^{\mu}),
\nonumber
\end{eqnarray}
\begin{eqnarray}
l^{\mu} &=& \left({2(r^2+a^2)\over \Delta}, ~~1, ~~0, ~~{2a\over \Delta}\right), \nonumber \\
n^{\mu} &=& \left(0, ~~{-\Delta \over 2\Sigma}, ~~0, ~~0\right), \\
m^{\mu} &=& {1\over \sqrt{2}(r+ia \cos \theta)}\left(ia\sin \theta, ~~0, ~~1,
~~{i\over \sin \theta}\right). \nonumber
\end{eqnarray}
Although it is expressed in these well-behaved ingoing Kerr coordinates, this null tetrad
is still singular at the horizon where $\Delta = 0$. At this point, notice that we can
get around this difficulty using the tetrad transformations. Namely, recall that the
orthogonality relations for null tetrad given in eq.(3) remain invariant under the
6-parameter group of homogeneous Lorentz transformations at each point of spacetime.
And this Lorentz group can be decomposed into 3-Abelian subgroups ;
\begin{eqnarray}
(I) ~~l&\rightarrow& l, ~~~m\rightarrow m+al,
    ~~~n\rightarrow n+a\bar{m}+\bar{a}m+a\bar{a}l, \nonumber \\
(II) ~~n&\rightarrow& n, ~~~m\rightarrow m+bn, 
    ~~~l\rightarrow l+b\bar{m}+\bar{b}m+b\bar{b}n,   \\
(III) ~~l&\rightarrow& \Lambda l, ~~~n\rightarrow \Lambda^{-1} n,
      ~~~m\rightarrow  e^{i\theta}m \nonumber
\end{eqnarray}
where $a$ and $b$ are complex numbers and $\Lambda$ and $\theta$ are real.
Each of these group transformations are called a ``null rotation'' and here
we particularly consider the null rotation (III). Under this null rotation (III),
the corresponding orthonormal tetrad $e_{A}$ is boosted in the $e_{1}=e_{\hat{r}}$
direction with 3-velocity ${(\Lambda^2-1)/(\Lambda^2+1)}$ and spatially rotated
about $e_{1}=e_{\hat{r}}$ through the angle $\theta$. Indeed this action is 
precisely what we need. Namely, in order to get a null tetrad well-behaved at
the horizon, we need to {\it boost it by an amount that becomes suitably infinite
on the horizon.} Thus we perform the null rotation (III) on the Kinnersley's
null tetrad given in ingoing null tetrad given above with 
$\Lambda = \Delta/2(r^2+a^2)$ and $e^{i\theta}=\Sigma^{1/2}/(r-ia\cos\theta)$
to obtain the following non-singular null tetrad on the horizon ;
\begin{eqnarray}
l^{\mu} &=& \left(1, ~~{\Delta \over 2(r^2+a^2)}, ~~0, ~~{a\over (r^2+a^2)}\right), 
\nonumber \\
n^{\mu} &=& \left(0, ~~{-(r^2+a^2)\over \Sigma}, ~~0, ~~0\right), \\
m^{\mu} &=& {1\over \sqrt{2}\Sigma^{1/2}}\left(ia\sin \theta, ~~0, ~~1,
~~{i\over \sin \theta}\right). \nonumber
\end{eqnarray}
This is the Hawking-Hartle (or Teukolsky) null tetrad and the associated covariant
components are given by
\begin{eqnarray}
l_{\mu} &=& \left({-\Delta \over 2(r^2+a^2)}, ~~{\Sigma\over (r^2+a^2)} ~~0, 
~~{\Delta\over 2(r^2+a^2)}a\sin^2 \theta \right),
\nonumber \\
n_{\mu} &=& \left({-(r^2+a^2)\over \Sigma}, ~~0, ~~0, ~~{(r^2+a^2)\over \Sigma}a\sin^2 \theta 
\right), \\
m_{\mu} &=& {1\over \sqrt{2}\Sigma^{1/2}}\left(-ia\sin \theta, ~~0, ~~\Sigma,
~~i(r^2+a^2)\sin \theta \right). \nonumber
\end{eqnarray}

{\rm \bf 2. Damour's quasi-orthonormal tetrad}

It is interesting to note that generally one can ``mix'' half of the null tetrad
$Z_{A}=(l,~n,~m,~\bar{m})$ and half of the orthonormal tetrad $e_{A}=(e_{0},~e_{1},
~e_{2},~e_{3})$ to form a ``quasi-orthonormal'' or ``mixed'' tetrad
\begin{eqnarray}
\left\{l^{\mu}, ~~-n^{\mu}, ~~e^{\mu}_{2}, ~~e^{\mu}_{3}\right\},
~~~\left\{-n_{\mu}, ~~l_{\mu}, ~~e^{2}_{\mu}, ~~e^{3}_{\mu}\right\}.
\end{eqnarray}
And if we construct this half-null, half-orthonormal, mixed tetrad from the previous
Hawking-Hartle null tetrad, it becomes Damour's quasi-orthonormal tetrad as we can see
shortly. Before we proceed, let us elaborate on the general construction of this
mixed tetrad. Using the relations between the orthonormal tetrad $e_{A}$ and null 
tetrad $Z_{A}$ given in eq.(2), 
\begin{eqnarray}
ds^2 &=& g_{\mu\nu}dx^{\mu}dx^{\nu} = \eta_{AB}e^{A}_{\mu}e^{B}_{\nu}dx^{\mu}dx^{\nu} 
\nonumber \\
&=& \left(-l_{\mu}n_{\nu} -n_{\mu}l_{\nu} +m_{\mu}\bar{m}_{\nu} +\bar{m}_{\mu}m_{\nu}\right)
dx^{\mu}dx^{\nu}
\end{eqnarray}
and hence $g_{\mu\nu} = -l_{\mu}n_{\nu} -n_{\mu}l_{\nu} +m_{\mu}\bar{m}_{\nu} +\bar{m}_{\mu}
m_{\nu}$. However, since the pair $(e_{0},~e_{1})$ is related only to $(l, ~n)$ while
the pair $(e_{2},~e_{3})$ is related only to $(m, ~\bar{m})$, one can write,
using $m_{\mu}\bar{m}_{\nu} +\bar{m}_{\mu}m_{\nu}=e^{2}_{\mu}e^{2}_{\nu}+e^{3}_{\mu}e^{3}_{\nu}$,
\begin{eqnarray}
g_{\mu\nu} &=& -l_{\mu}n_{\nu} -n_{\mu}l_{\nu} +e^{2}_{\mu}e^{2}_{\nu} +e^{3}_{\mu}e^{3}_{\nu}
~~~{\rm or} \nonumber \\
g^{\mu\nu} &=& -l^{\mu}n^{\nu} -n^{\mu}l^{\nu} +e^{\mu}_{2}e^{\nu}_{2} +e^{\mu}_{3}e^{\nu}_{3}.
\end{eqnarray}
This obviously implies that one may mix half of null tetrad and half of orthonormal tetrad
to form a mixed tetrad as given in eq.(12). Therefore, we now construct this mixed tetrad
from the previous Hawking-Hartle tetrad as
\begin{eqnarray}
e^{\mu}_{0} &\equiv& l^{\mu} = \left(1, ~~{\Delta \over 2(r^2+a^2)}, ~~0, 
~~{a\over (r^2+a^2)}\right), \nonumber \\
e^{\mu}_{1} &\equiv& -n^{\mu} = \left(0, ~~{(r^2+a^2)\over \Sigma}, ~~0, ~~0\right), \\
e^{\mu}_{2} &=& {1\over \sqrt{2}}(m^{\mu}+\bar{m}^{\mu}) = \left(0, ~~0, 
~~{1\over \Sigma^{1/2}}, ~~0\right) \nonumber \\
e^{\mu}_{3} &=& {1\over \sqrt{2i}}(m^{\mu}-\bar{m}^{\mu}) = \left({a\sin \theta \over \Sigma^{1/2}},
~~0, ~~0, ~~{1\over \Sigma^{1/2}\sin \theta}\right) \nonumber
\end{eqnarray}
and its dual is
\begin{eqnarray}
e^{0}_{\mu} &\equiv& -n_{\mu} = \left({(r^2+a^2)\over \Sigma}, ~~0, ~~0,
~~{-(r^2+a^2)\over \Sigma}a\sin \theta \right), \nonumber \\
e^{1}_{\mu} &\equiv& l_{\mu} = \left({-\Delta \over 2(r^2+a^2)}, ~~{\Sigma \over (r^2+a^2)}, ~~0,
~~{\Delta \over 2(r^2+a^2)}a\sin^2 \theta \right), \\
e^{2}_{\mu} &=& {1\over \sqrt{2}}(m_{\mu}+\bar{m}_{\mu}) = \left(0, ~~0,
~~\Sigma^{1/2}, ~~0\right) \nonumber \\
e^{3}_{\mu} &=& {1\over \sqrt{2i}}(m_{\mu}-\bar{m}_{\mu}) = \left({-a\sin \theta \over \Sigma^{1/2}},
~~0, ~~0, ~~{(r^2+a^2)\over \Sigma^{1/2}}\sin \theta \right) \nonumber
\end{eqnarray}
where we renamed as 
\begin{eqnarray}
l^{\mu} &\rightarrow& e^{\mu}_{0}, ~~~n^{\mu} \rightarrow -e^{\mu}_{1}, \nonumber \\
l_{\mu} &\rightarrow& e^{1}_{\mu}, ~~~n_{\mu} \rightarrow -e^{0}_{\mu}, \nonumber
\end{eqnarray}
to go from the null tetrad's orthogonality relations
$-l^{\mu}n_{\mu} = 1 = m^{\mu}\bar{m}_{\mu}$ to the usual orthonormality condition
$e^{\mu}_{A}e^{B}_{\mu}=\delta_{A}^{B}$, $e^{\mu}_{A}e^{A}_{\nu}=\delta^{\mu}_{\nu}$.
Note that this mixed tetrad precisely coincides with Damour's choice of quasi-orthonormal
tetrad [6]. And the tetrad metric $\epsilon_{AB} = \epsilon^{AB}$ in
\begin{eqnarray}
ds^2 = \epsilon_{AB}e^{A}e^{B} \nonumber
\end{eqnarray}
can be identified with
\begin{eqnarray}
\epsilon_{AB} = \epsilon^{AB} =
\pmatrix{0 & 1 & 0 & 0 \cr
         1 & 0 & 0 & 0 \cr
         0 & 0 & 1 & 0 \cr
         0 & 0 & 0 & 1 \cr}. \nonumber
\end{eqnarray}
Particularly, for later use, we explicitly write down the dual basis of vectors as
$e_{A} = \left(e_{0}=e_{\hat{v}}, ~e_{1}=e_{\hat{r}}, ~e_{2}=e_{\hat{\theta}},
~e_{3}=e_{\hat{\phi}}\right)$,
\begin{eqnarray}
e_{0} &=& \partial_{v} + {\Delta\over 2(r^2+a^2)}\partial_{r} + {a\over (r^2+a^2)}\partial_{\phi},
\nonumber \\
e_{1} &=& {(r^2+a^2)\over \Sigma}\partial_{r}, \nonumber \\
e_{2} &=& {1\over \Sigma^{1/2}}\partial_{\theta}, \\
e_{3} &=& {1\over \Sigma^{1/2}}\left[a\sin \theta  \partial_{v} + {1\over \sin \theta}
\partial_{\phi}\right]. \nonumber
\end{eqnarray}
Note that in all the calculations involved in this work to read off physical components
of tensors such as Maxwell field tensor and current 4-vector, we shall strictly use
this quasi-orthonormal tetrad given in eqs.(15) and (16) and nothing else. In this sense,
our choice of local tetrad frame is slightly different from that in the original work of
Damour [6] in which he introduced, particularly on the 2-dimensional, $v=$const. section of
the event horizon, some other orthonormal basis (slightly different from $\{e^{\mu}_{2},
~e^{\mu}_{3}\}$ given above) specially adapted to the ``intrinsic geometry'' of the 
$v=const.$ section of the horizon and used them to project out physical components of tensors. \\ 
Before we proceed, we momentarily recall the ``Zero-Angular-Momentum-Observer (ZAMO)''
tetrad in Boyer-Lindquist coordinates for the sake of comparison with this
quasi-orthonormal tetrad in ingoing Kerr coordinates. The dual of ZAMO tetrad is
given by
$\tilde{e}_{A} = \left(\tilde{e}_{0}=\tilde{e}_{(t)}, ~\tilde{e}_{1}=\tilde{e}_{(r)},
~\tilde{e}_{2}=\tilde{e}_{(\theta)}, ~\tilde{e}_{3}=\tilde{e}_{(\tilde{\phi})}\right)$,
\begin{eqnarray}
\tilde{e}_{0} &=& {1\over \alpha}[\partial_{t} + \Omega \partial_{\tilde{\phi}}],
\nonumber \\
\tilde{e}_{1} &=& \left({\Delta \over \Sigma}\right)^{1/2}\partial_{r}, \nonumber \\
\tilde{e}_{2} &=& {1\over \Sigma^{1/2}}\partial_{\theta}, \\
\tilde{e}_{3} &=& \left({\Sigma \over A}\right)^{1/2}{1\over \sin \theta } \partial_{\tilde{\phi}}
\nonumber
\end{eqnarray}
where 
$\alpha^2 = \left(\Sigma \Delta / A \right)$,
$A = [(r^2+a^2)^2 - \Delta a^2 \sin^2 \theta]$,
$\Omega = -g_{t\tilde{\phi}}/g_{\tilde{\phi}\tilde{\phi}} = 2Mra/A $
ith $\alpha $ being the lapse function. 
As is well-known, this ZAMO tetrad, particularly
$\tilde{e}_{(t)}$ exhibits pathological behavior as the horizon is approached, i.e., in the
limit, $\Delta \rightarrow 0$ or $\alpha \rightarrow 0$ and it can be attributed to the
Boyer-Lindquist coordinate system itself as it is ill-defined at the horizon. As a matter of
fact, this is precisely the reason why we (and originally Damour) choose to work in ingoing
Kerr coordinates and further employ half-null, half-orthogonal tetrad instead despite its
seemingly complex structure. ZAMO is a ``FIDO'' (fiducial observer) and 
$\tilde{e}^{\mu}_{(t)}=u^{\mu}$ (in $\tilde{e}_{(t)}=\tilde{e}^{\mu}_{(t)}\partial_{\mu}$)
is its 4-velocity whose pathological behavior near the horizon needs to be regularized,
for instance, by $\tilde{e}^{\mu}_{(t)}\rightarrow \alpha \tilde{e}^{\mu}_{(t)}=
(\partial/\partial t)^{\mu} + \Omega (\partial/\partial \tilde{\phi})^{\mu}$. Obviously,
this regularized 4-velocity of ZAMO becomes, at the horizon, Killing vector normal to the 
horizon, $\tilde{\chi}^{\mu} = (\partial/\partial t)^{\mu} + \Omega_{H} 
(\partial/\partial \tilde{\phi})^{\mu}$ (with $\Omega_{H}=a/(r_{+}^2 + a^2)$ being the
angular velocity of the horizon) and hence is null. Only after this regularization, the
(dual) of ZAMO tetrad is now made to be well-defined at the future horizon and then can be 
used, say, to read out physical components of a given tensor via eq.(5). \\
We now consider this standard procedure toward the study of electrodynamics in the
background of Kerr black hole geometry by employing, instead, Damour's quasi-orthonormal
tetrad in ingoing Kerr coordinates. The first thing that can be noticed is the fact that 
the 4-velocity of a local observer in this quasi-orthonormal frame, $e^{\mu}_{\hat{v}}$
(in $e_{\hat{v}}=e^{\mu}_{\hat{v}}\partial_{\mu}$) becomes, at the horizon where
$\Delta = 0$, at once, the usual Killing vector normal to the horizon,
$\chi^{\mu} = (\partial/\partial v)^{\mu} + \Omega_{H}(\partial/\partial \phi)^{\mu}$
which has no pathological behavior whatsoever. Thus we do not need any {\it ad hoc}
regularization prescription. As we have carefully discussed earlier in the derivation
of Hawking-Hartle (or Teukolsky) tetrad, it is interesting to note that this regular
behavior of the 4-velocity at the horizon is due to neither the choice of ingoing
Kerr coordinates (which is known to be well-behaved at the horizon) nor the employment
of the (half) null nature of the tetrad but really due to the action of ``null
rotation III'' in which particularly the associated orthonormal tetrad is boosted
in the $e_{\hat{r}}$-direction with 3-velocity $(\Lambda^2-1)/(\Lambda^2+1)$ with
$\Lambda = \Delta/2(r^2+a^2)$. Namely the lesson there was that simply taking null
tetrad is not enough and in order to get a well-behaved null tetrad at the horizon,
one needs to boost it by an amount that becomes suitably infinite on the horizon.
Having a well-behaved tetrad (at the horizon) in our possession, we now can proceed
and calculate ``physical'' components of any given tensor by projecting its components
onto this quasi-orthonormal tetrad frame. To summarize, in this comparison between
the choice of ZAMO in Boyer-Lindquist coordinates and that of Damour's quasi-
orthonormal tetrad in ingoing Kerr coordinates, it appears that the latter is physically
more relevant in that it has been constructed in a more natural manner than the former
which, for regularity at the horizon, involves somewhat {\it ad hoc} prescription.
Thus in the present work, we choose to work with Damour's quasi-orthonormal tetrad in
ingoing Kerr coordinates and try to read out physical components of all tensors involved.
For example, physical components of Maxwell field strength and electric current 4-vector
will be identified with
$F_{AB} = F_{\mu\nu}(e^{\mu}_{A}e^{\nu}_{B})$, $J^{A} = J^{\mu}e^{A}_{\mu}$
respectively.

\begin{center}
{\rm \bf III. Identification of electric and magnetic fields on the horizon}
\end{center}

As stated, our major concern in the present work is the derivation of curious boundary
conditions for electromagnetic fields at the horizon but in the context of non-linear
BI electrodynamics. As we shall see in a moment, the highly non-linear BI equations can 
be made to take on a seemingly linear structure similar to that of Maxwell equations.
And to this end, we need to introduce two species of field strength tensors ; the new
one $G_{\mu\nu}$ for the inhomogeneous BI field equations and the usual one $F_{\mu\nu}$ for
the homogeneous Bianchi identity. Despite this added technical complexity, however, the
basic field quantities, namely the physical (finite and non-zero) components of electric 
field and magnetic induction can still be extracted from the standard field strength
$F_{\mu\nu}$. Thus before we go on, it might be relevant to remind two alternative typical
procedures by which one can read off physical components of electric field and magnetic 
induction from $F_{\mu\nu}$, generally. Obviously, the first procedure involves the
projection of components of  $F_{\mu\nu}$ onto the orthonormal tetrad frame chosen,
$F_{AB} = F_{\mu\nu}(e^{\mu}_{A}e^{\nu}_{B})$ as given above. Since $A$,$B$ are now
tangent space indices in this locally-flat tetrad frame, the physical electric and 
magnetic field components then can be read off in a standard manner as 
\begin{eqnarray}
F_{AB} = \{F_{i0}, F_{ij}\} \nonumber
\end{eqnarray}
where 
\begin{eqnarray}
E_{i} &=& F_{i0}, \\
B_{i} &=& {1\over 2}\epsilon_{ijk}F^{jk} = {1\over 2}\epsilon_{0ijk}F^{jk} = 
\tilde{F}_{0i}= - \tilde{F}_{i0}. \nonumber
\end{eqnarray}
The second alternative but equivalent procedure can be described as follows. Consider
a family of fiducial observers (FIDOs) whose worldlines are a congruence of timelike
curves orthogonal to spacelike hypersurfaces. Let $u^{\mu}$ be the 4-velocity of a
FIDO normalized as $u^{\alpha}u_{\alpha} = -1$. Since, by definition, all the physically
meaningful measurements should be made by these FIDOs, one can expect that the local 
values of the electric and magnetic fields measured by a FIDO with 4-velocity would be
given by
\begin{eqnarray}
E^{\alpha} &=& F^{\alpha \beta}u_{\beta}, \\
B^{\alpha} &=& -{1\over 2}\epsilon^{\alpha \beta \lambda \sigma}u_{\beta}F_{\lambda \sigma}
= - \tilde{F}^{\alpha \beta}u_{\beta} \nonumber
\end{eqnarray}
or
\begin{eqnarray}
F^{\alpha \beta} = u^{\alpha}E^{\beta}-E^{\alpha}u^{\beta}+\epsilon^{\alpha \beta
\lambda \sigma}u_{\lambda}B_{\sigma}. \nonumber
\end{eqnarray}
In addition, since both the electric and the magnetic fields are purely spatial vectors,
one may expect
\begin{eqnarray}
u_{\alpha}E^{\alpha} = 0 = u_{\alpha}B^{\alpha}. 
\end{eqnarray}
Finally, we project $E^{\mu}=(0,E^{i})$ and $B^{\mu}=(0,B^{i})$ onto the orthonormal tetrad frame 
associated with this FIDO to get their physical (finite and non-zero) components
\begin{eqnarray}
E^{A} = e^{A}_{\mu}E^{\mu},  ~~~B^{A} = e^{A}_{\mu}B^{\mu}
\end{eqnarray}
then $E_{i} = E^{i}$, $B_{i} = B^{i}$ where $A=(0,i)$. Thus in order to evaluate physical
components of electric field and magnetic induction $(E_{i}, B_{i})$, one may choose
between these two procedures and in the present work, where the quasi-orthonormal tetrad
of Damour is already available, we shall employ the first procedure for actual calculations. 

{\rm \bf 1. Brief review of BI electrodynamics in curved spacetimes}

Eventually for the exploration of boundary conditions for BI electromagnetic fields at the
horizon of Kerr hole, we now briefly describe general formulation of BI theory in a given
curved spacetime. Since this BI theory of electromagnetism is, despite its long history and
physically interesting motivations behind it, not well-known and hence might be rather 
unfamiliar to relativists and workers in theoretical astrophysics community, we provide
an introductory review of BI theory in flat spacetime in Appendix. For a recent study
of this flat spacetime BI theory particularly in modern field theory perspective, we
also refer the reader to [15].
In our discussion below, however, we are implicitly aimed at adapting the theory
to the formulation of electrodynamics in a rotating uncharged black hole spacetime.
Also at this point in seems worthy of mention that throughout, we will be assuming the
``weak field limit''. To be a little more concrete, we consider the dynamics of 
electromagnetic field governed by the BI theory in the background of uncharged Kerr
black hole spacetime. And we assume that the strength of this external electromagnetic
field is small enough not to have any sizable backreaction to the background geometry.
Then this means we are not considering a phenomena described by solutions in coupled
full Einstein-BI theory but an environment where the test electromagnetic field
possesses dynamics governed by the BI theory rather than by the Maxwell theory. 
Also note that this assumption can be further justified as long as we confine our
concern to the electrodynamics around the ``uncharged'' Kerr black hole. If, instead,
one is interested in the same physics but in charged rotating black holes (which,
however, is rather uninteresting since it is less likely to happen in realistic
astrophysical environments where the black hole charge, if any, gets quickly
neutralized by the surrounding plasma), one would have to deal with the full
Einstein-BI theory in which, unfortunately, the charged rotating black hole solution
is not available. \\
Thus we consider here the action of (4-dimensional) BI theory in a fixed background
spacetime with metric $g_{\mu\nu}$. And to do so, some explanatory comments might
be relevant. Coupling the BI gauge theory to gravity is not so familiar and hence
we start first with the BI theory action in 4-dim. flat spacetime.
\begin{eqnarray}
S &=& \int d^4x {1\over 4\pi}\beta^2  \left[1 - \sqrt{-det\left(\eta_{\mu\nu}+
{1\over \beta}F_{\mu\nu}\right)}\right] \nonumber \\
&=&  \int d^4x {1\over 4\pi}\beta^2 \left[1 - \sqrt{1+{1\over 2\beta^2}F_{\mu\nu}
F^{\mu\nu}-{1\over 16\beta^4}\left(F_{\mu\nu}\tilde{F}^{\mu\nu}\right)^2}\right] 
\nonumber
\end{eqnarray}
and then elevate it to its curved spacetime version by employing the minimal
coupling scheme. This is really the conventional procedure and the result is
\begin{eqnarray}
S = \int d^4x\sqrt{g} \left\{{1\over 4\pi}\beta^2 \left[1 - \sqrt{1+{1\over
2\beta^2}\left(g^{\mu\alpha}g^{\nu\beta}F_{\mu\nu}F_{\alpha\beta}\right)
-{1\over 16\beta^4}\left(g^{\mu\alpha}g^{\nu\beta}F_{\mu\nu}
\tilde{F}_{\alpha\beta}\right)^2}\right]+J^{\mu}A_{\mu}\right\}
\end{eqnarray}
where $J^{\mu}=\rho_{e}u^{\mu}+j^{\mu}_{e}$ is the electric source current for
the vector potential $A_{\mu}$. Here, the generic parameter of the theory
``$\beta$'' having the canonical dimension $dim[\beta]=dim[F_{\mu\nu}]=+2$,
probes the degree of deviation of BI theory from the standard Maxwell theory
as the limit $\beta \rightarrow \infty$ obviously corresponds to the
Maxwell theory action. Now extremizing this action with respect to $A_{\mu}$ 
yields the dynamical BI field equation
\begin{eqnarray}
\nabla_{\nu}\left[{F^{\mu\nu}-{1\over 4\beta^2}(F_{\alpha\beta}
\tilde{F}^{\alpha\beta})\tilde{F}^{\mu\nu} \over \sqrt{1+{1\over 2\beta^2}
(F_{\alpha\beta}F^{\alpha\beta})-{1\over 16\beta^4}(F_{\alpha\beta}
\tilde{F}^{\alpha\beta})^2}}\right] = 4\pi J^{\mu}
\end{eqnarray}
while the Bianchi identity, which is a supplementary equation to this field
equation is given by
\begin{eqnarray}
\nabla_{\nu}\tilde{F}^{\mu\nu} = {1\over \sqrt{g}}\partial_{\nu}\left[
\sqrt{g}\tilde{F}^{\mu\nu}\right] = 0
\end{eqnarray}
where $\tilde{F}^{\mu\nu}={1\over 2}\epsilon^{\mu\nu\alpha\beta}F_{\alpha\beta}$
is the Hodge dual of $F_{\mu\nu}$. Note that this Bianchi identity is just
a geometrical equation independent of the detailed nature of a gauge theory 
action. Thus it remains the same as that in Maxwell theory.
For later use, we also provide the energy-momentum tensor of this BI theory,
\begin{eqnarray}
T_{\mu\nu} &=& {2\over \sqrt{g}}{\delta S\over \delta g^{\mu\nu}} \\
&=& {1\over 4\pi}\left\{\beta^2 (1-R) g_{\mu\nu} + {1\over R}\left[
F_{\mu\alpha}F_{\nu}^{\alpha}-{1\over 4\beta^2}(F_{\alpha\beta}
\tilde{F}^{\alpha\beta})F_{\mu\alpha}\tilde{F}_{\nu}^{\alpha}\right]\right\}
\nonumber
\end{eqnarray}
where $R\equiv \left[1+{1\over 2\beta^2}
(F_{\alpha\beta}F^{\alpha\beta})-{1\over 16\beta^4}(F_{\alpha\beta}
\tilde{F}^{\alpha\beta})^2\right]^{1/2}$.
Now the first thing that one can readily notice in this rather unfamiliar BI
theory of electrodynamics might be the fact that even in the absence of the
source current, the dynamical BI field equation and the geometrical Bianchi
identity clearly are not dual to each other under $F_{\mu\nu}\rightarrow
\tilde{F}_{\mu\nu}$ and $\tilde{F}_{\mu\nu}\rightarrow - F_{\mu\nu}$.
Obviously, this is in contrast to what happens in the standard Maxwell theory
and can be attributed to the fact that when passing from the Maxwell to this
highly non-linear BI theory, only the dynamical field equation undergoes
non-trivial change (``non-linearization'') and the geometrical Bianchi
identity, as pointed out above, remains unchanged. Therefore in order to
deal with this added complexity properly and formulate the BI theory in
curved background spacetime in a manner parallel to that for the standard 
Maxwell theory, we find it relevant to introduce another field strength
$G_{\mu\nu}$ which, however, is made up of $F_{\mu\nu}$ and $\tilde{F}_{\mu\nu}$.
To be more precise, consider introducing, for the inhomogeneous BI field
equation,
\begin{eqnarray}
G_{\mu\nu} = {1\over R}\left[F_{\mu\nu} - {1\over 4\beta^2}(F_{\alpha\beta}
\tilde{F}^{\alpha\beta})\tilde{F}_{\mu\nu}\right]
\end{eqnarray}
and defining the associated fields on each spacelike hypersurfaces,
$(D^{\alpha}, H^{\alpha})$ as
\begin{eqnarray}
D^{\alpha} &=& G^{\alpha\beta}u_{\beta}, \\
H^{\alpha} &=& -{1\over 2}\epsilon^{\alpha\beta\lambda\sigma}u_{\beta}
G_{\lambda\sigma} = - \tilde{G}^{\alpha\beta}u_{\beta} \nonumber
\end{eqnarray}
which also implies their purely spatial nature
\begin{eqnarray}
u_{\alpha}D^{\alpha} = 0 = u_{\alpha}H^{\alpha}.
\end{eqnarray}
As before, $u^{\mu}$ here is the 4-velocity of FIDO (or more precisely ZAMO
for rotating Kerr geometry) having a timelike geodesic orthogonal to spacelike
hypersurfaces. Then the inhomogeneous BI field equation now takes the form
\begin{eqnarray}
\nabla_{\nu}G^{\mu\nu} = 4\pi J^{\mu}
\end{eqnarray}
which relates the fields $(D^{\mu}, H^{\mu})$ as defined above to ``free''
charge and current $J^{\mu}=\rho_{e}u^{\mu}+j_{e}^{\mu}$.
Despite this extra elaboration, the fundamental field quantities, namely
the electric field and the magnetic induction still can be identified with
\begin{eqnarray}
E^{\alpha} &=& F^{\alpha\beta}u_{\beta}, \\
B^{\alpha} &=& -{1\over 2}\epsilon^{\alpha\beta\lambda\sigma}u_{\beta}
F_{\lambda\sigma} = - \tilde{F}^{\alpha\beta}u_{\beta} \nonumber
\end{eqnarray} 
which, as before, implies
$u_{\alpha}E^{\alpha} = 0 = u_{\alpha}B^{\alpha}$. Thus the homogeneous
Bianchi identity equation
\begin{eqnarray}
\nabla_{\nu}\tilde{F}^{\mu\nu} = 0
\end{eqnarray}
is expressible in terms of usual $(E^{\mu},B^{\mu})$ fields. Then in this
new representation of a set of BI equations, we now imagine their
space-plus-time decomposition. Obviously, the dynamical BI field equation
would split up into two inhomogeneous equations involving
$(D^{\mu},H^{\mu})$ and the ``free'' source charge and current
$J^{\mu}=\rho_{e}u^{\mu}+j_{e}^{\mu}$ whereas the geometrical Bianchi 
identity equation decomposes into two homogeneous equations involving
$(E^{\mu},B^{\mu})$. Incidentally, one can then realize that this 
indeed is reminiscent of Maxwell equations in a ``medium''. Namely,
in this new representation, the BI theory of electrodynamics can be
thought of as taking on the structure of ordinary Maxwell electrodynamics
in a medium with non-trivial electric susceptibility and magnetic
permeability. In this interpretation of the new representation of the
BI theory, then, it is evident that the system is of course not linear
in that $(D^{\mu},H^{\mu})$ and $(E^{\mu},B^{\mu})$ are related by
\begin{eqnarray}
D^{\mu} &=& {1\over R}\left[E^{\mu} + {1\over \beta^2}(E_{\alpha}
B^{\alpha})B^{\mu}\right], \nonumber \\
H^{\mu} &=& {1\over R}\left[B^{\mu} - {1\over \beta^2}(E_{\alpha}
B^{\alpha})E^{\mu}\right], \\
&&{\rm with} ~~~R = \left[1 - {1\over \beta^2}(E_{\alpha}E^{\alpha}-
B_{\alpha}B^{\alpha}) - {1\over \beta^4}(E_{\alpha}B^{\alpha})^2
\right]^{1/2} \nonumber
\end{eqnarray}
or inversely
\begin{eqnarray}
E^{\mu} &=& {1\over R}\left[D^{\mu} - {1\over \beta^2}(D_{\alpha}
H^{\alpha})H^{\mu}\right], \nonumber \\
B^{\mu} &=& {1\over R}\left[H^{\mu} + {1\over \beta^2}(D_{\alpha}
H^{\alpha})D^{\mu}\right], \\
&&{\rm with} ~~~R = \left[1 - {1\over \beta^2}(H_{\alpha}H^{\alpha}-
D_{\alpha}D^{\alpha}) - {1\over \beta^4}(D_{\alpha}H^{\alpha})^2
\right]^{1/2} \nonumber
\end{eqnarray}
where we usued eqs.(27),(28) and (31) and $u^{\alpha}u_{\alpha}=-1$,
$F_{\alpha\beta}F^{\alpha\beta}=-2(E_{\alpha}E^{\alpha}-
B_{\alpha}B^{\alpha})$ and $F_{\alpha\beta}\tilde{F}^{\alpha\beta}=
4E_{\alpha}B^{\alpha}$. It is also noteworthy from above expressions
that
\begin{eqnarray}
E_{\alpha}B^{\alpha} = D_{\alpha}H^{\alpha}.
\end{eqnarray}
Thus from now on, we may call $D^{\mu}=(0, D^{i})$ as the ``electric
displacement'' 4-vector and $H^{\mu}=(0, H^{i})$ as the ``magnetic 
field strength'' 4-vector.

{\rm \bf 2. Electric field and magnetic induction on the horizon}

Earlier, we mentioned that we shall employ, between the two alternative
procedures to evaluate ``physical'' components of electric field and
magnetic induction, the first one. In the context of BI theory of
electrodynamics, however, there are a set of fields $D^{\mu}=(0, D^{i}=D_{i})$,
$H^{\mu}=(0, H^{i}=H_{i})$ in addition to $E^{\mu}=(0, E^{i}=E_{i})$,
$B^{\mu}=(0, B^{i}=B_{i})$. Then we shall first evaluate $(D_{i}, H_{i})$
on the horizon and then from them identify  $(E_{i}, B_{i})$ afterwards.
With respect to Damour's quasi-orthonormal tetrad, the physical components
of electric displacement $D_{i}$ and magnetic field strength $H_{i}$ can be
read off as
\begin{eqnarray}
G_{AB} &=& G_{\mu\nu}(e^{\mu}_{A}e^{\nu}_{B}) ~~~~{\rm and} \nonumber \\
D_{i} &=& G_{i0},
~~~~H_{i} = {1\over 2}\epsilon_{ijk}G^{jk} = - \tilde{G}_{i0}.
\end{eqnarray}
More concretely, since we are working in ingoing Kerr coordinates
$(v, r, \theta, \phi)$, the components of electric displacement {\it on the horizon}
can be read off as 
\begin{eqnarray}
D_{\hat{r}} &=& D_{1} = G_{10} = G_{\mu\nu}(e^{\mu}_{1}e^{\nu}_{0})|_{r_{+}} \nonumber \\
&=& {(r^2_{+}+a^2)\over \Sigma_{+}}\left[G_{rv}+{a\over (r^2_{+}+a^2)}G_{r\phi}\right], 
\nonumber \\
D_{\hat{\theta}} &=& D_{2} = G_{20} = G_{\mu\nu}(e^{\mu}_{2}e^{\nu}_{0})|_{r_{+}} \nonumber \\
&=& {1\over \Sigma^{1/2}_{+}}\left[G_{\theta v}+{a\over (r^2_{+}+a^2)}G_{\theta \phi}\right],\\
D_{\hat{\phi}} &=& D_{3} = G_{30} = G_{\mu\nu}(e^{\mu}_{3}e^{\nu}_{0})|_{r_{+}} \nonumber \\
&=& {\Sigma^{1/2}_{+}\over (r^2_{+}+a^2)\sin \theta}G_{\phi v} \nonumber
\end{eqnarray}
where $\Sigma_{+}\equiv (r^2_{+}+a^2\cos^2 \theta)$. Next, the components of magnetic field
strength again on the horizon can be read off as
\begin{eqnarray}
H_{\hat{r}} &=& H_{1} = -\tilde{G}_{10} = G^{23} = G_{23} \nonumber \\
 &=& G_{\mu\nu}(e^{\mu}_{2}e^{\nu}_{3})|_{r_{+}} 
 = {1\over \Sigma_{+}\sin \theta}[a\sin^2 \theta G_{\theta v} + G_{\theta \phi}], \nonumber \\  
H_{\hat{\theta}} &=& H_{2} = -\tilde{G}_{20} = G^{31} = G_{30} \\
 &=& D_{\hat{\phi}} = {\Sigma^{1/2}_{+}\over (r^2_{+}+a^2)\sin \theta}G_{\phi v}, \nonumber \\
H_{\hat{\phi}} &=& H_{3} = -\tilde{G}_{30} = G^{12} = G_{02} \nonumber \\
 &=& -D_{\hat{\theta}} = - {1\over \Sigma^{1/2}_{+}}\left[G_{\theta v}+{a\over
(r^2_{+}+a^2)}G_{\theta \phi}\right] \nonumber 
\end{eqnarray}
where we used the Damour's quasi-orthonormal tetrad metric
\begin{eqnarray}
ds^2 = 2e^{0}e^{1} + e^{2}e^{2} + e^{3}e^{3} = \epsilon_{AB}e^{A}e^{B} \nonumber
\end{eqnarray}
to deduce
\begin{eqnarray}
G^{23} &=& \epsilon^{2A}\epsilon^{3B}G_{AB} = G_{23}, \nonumber \\
G^{31} &=& \epsilon^{3A}\epsilon^{1B}G_{AB} = G_{30}, \nonumber \\
G^{12} &=& \epsilon^{1A}\epsilon^{2B}G_{AB} = G_{02}. \nonumber
\end{eqnarray}
Thus it is interesting to note that on the horizon
$H_{\hat{\theta}} = D_{\hat{\phi}}$ and $H_{\hat{\phi}} = - D_{\hat{\theta}}$ or
in a vector notation in a tangent space to the horizon,
\begin{eqnarray}
\vec{H}_{H} = \vec{D}_{H} \times \hat{n} 
\end{eqnarray}
where $\hat{n} = \hat{r}$ is the vector (outer) normal to the horizon. 
This relation indicates that $\{\vec{H}_{H}, ~\vec{D}_{H}, ~\hat{n}\}$ form a 
``triad'' on the horizon and hence constitutes the so-called ``radiative ingoing
(or, inward Poynting flux)'' boundary condition at horizon as seen by a local 
observer at rest in the quasi-orthonormal tetrad frame. Here, however, it seems worthy
of note that although this relation is one of the horizon boundary conditions 
eventually we are after, it has {\it not} been obtained essentially from the 
horizon specifics. As a matter of fact, it holds for any $r=const.$ sections 
and indeed its emergence can be attributed to the ``half-null'' ($e^{\mu}_{0}=l^{\mu}$,
~$e^{\mu}_{1}=-n^{\mu}$) structure of Damour's quasi-orthonormal tetrad.
Given the observation that the same type of relation as this ``radiative ingoing
boundary condition'' actually holds for any null surface, one might wonder what
then would be the distinctive nature of the event horizon (among null surfaces)
that actually endows this boundary condition with real physical meaning.
Znajek [5] provided one possible answer to this question and it is : the special
feature of the event horizon over all other null surfaces is that it is a ``stationary''
null surface and there is a natural class of time coordinates associated with the
frame at infinity in which the black hole is at rest. And the physical components
of electric and magnetic fields should be evaluated, in a unique way, in a frame
at rest on the horizon. At this point, we remark on another crucial thing happening
at the horizon. Namely we note that at the horizon,
\begin{eqnarray}
D_{\alpha}H^{\alpha} &=& g_{\alpha \beta}D^{\alpha}H^{\beta} = (\epsilon_{AB}
e^{A}_{\alpha}e^{B}_{\beta})D^{\alpha}H^{\beta} \nonumber \\
&=& \epsilon_{AB}D^{A}H^{B} = D^{0}H^{1}+D^{1}H^{0}+D^{2}H^{2}+D^{3}H^{3} =
D^{2}D^{3} - D^{3}D^{2} = 0, \nonumber \\
D_{\alpha}D^{\alpha} &=& g_{\alpha \beta}D^{\alpha}D^{\beta} = (\epsilon_{AB}    
e^{A}_{\alpha}e^{B}_{\beta})D^{\alpha}D^{\beta} \\
&=& \epsilon_{AB}D^{A}D^{B} = D^{0}D^{1}+D^{1}D^{0}+D^{2}D^{2}+D^{3}D^{3} =
D^{2}D^{2} + D^{3}D^{3} \nonumber \\
&=& H^{3}H^{3} + H^{2}H^{2} = \epsilon_{AB}H^{A}H^{B} = 
g_{\alpha \beta}H^{\alpha}H^{\beta} = H_{\alpha}H^{\alpha}. \nonumber 
\end{eqnarray}
These relations also holds {\it not} only at the horizon but on any 
$r=const.$ sections and again can be attributed to the half-null nature of
Damour's quasi-orthonormal tetrad. One immediate consequence of these relations
$D_{\alpha}H^{\alpha}=0$ and $D_{\alpha}D^{\alpha}= H_{\alpha}H^{\alpha}$
everywhere is that practically $E^{\mu}=D^{\mu}$ and $B^{\mu}=H^{\mu}$
everywhere (due to eqs. (33) and (34))
as seen by a local observer at rest in the quasi-orthonormal tetrad
frame. In fact, the interpretation of this is straightforward. Since Damour's
quasi-orthonormal tetrad is half-null in ($v-r$) sector, an observer in this
tetrad frame is actually a null observer who, as a result of his motion, would
see the electromagnetic field around him as a ``radiation field'' all the way
which, in turn, turns the BI theory of electrodynamics effectively into the
Maxwell theory. What is particularly remarkable concerning this study of 
electrodynamics in the background of Kerr black hole in the context of BI theory
is that the nature of the theory or the concrete structure of BI equations happens
to be such that it is indistinguishable to a local observer in Damour's
quasi-orthonormal tetrad frame (indeed to any null observers) from that of
standard Maxwell theory. This point is indeed quite amusing on theoretical side.
From now on, then, whenever we deal with quantities involving physical components
of fields as seen by this observer in Damour's tetrad frame, we can freely replace
($D^{\mu}(E^{\mu})$, ~$H^{\mu}(B^{\mu})$) by ($E^{\mu}(D^{\mu})$, ~$B^{\mu}(H^{\mu})$).
Thus the radiative ingoing boundary condition at the horizon obtained above can be
given in terms of electric field and magnetic induction as
\begin{eqnarray}
\vec{B}_{H} = \vec{E}_{H} \times \hat{n}. 
\end{eqnarray}
As pointed out earlier, this relation states that the electric and magnetic
fields tangential to the horizon are equal in magnitude and perpendicular in direction and
hence their Poynting energy flux is {\it into} the hole. This boundary condition as seen 
by a local observer again in a null tetrad frame (which has been made to be well-behaved
at the horizon by the amount of  boost that becomes suitably infinite at the horizon) has been
derived first by Znajek [5] in the context of standard Maxwell theory and here we just 
witnessed that precisely the same radiative ingoing boundary condition holds in the BI theory
context as well.  

\begin{center}
{\rm \bf IV. (Fictitious) Charge and current on the horizon}
\end{center}

It is well appreciated that in any attempt
to have an intuitive picture of Blandford-Znajek process for the rotational energy extraction
from rotating black holes, the introduction of surface charge and current density on the
(stretched) horizon proves to be quite convenient. For instance, the circuit analysis
in the membrane paradigm [11] cannot do without the notion of the horizon surface charge and
current density. If one follows the original argument of Damour [6], one can justify their
introduction as follows. Suppose the existence of a 4-current $J^{\mu}(v,r,\theta,\phi)$
which is defined and conserved all over the spacetime. Let $r=r_{+}$ be the location of
an event horizon, then obviously some charge and current can plunge into the hole and
disappear from the region $r>r_{+}$. Nevertheless, imagine that we do not want to consider
what happens inside the black hole ($r<r_{+}$) and just wish to keep the charge and
current conserved in the region $r>r_{+}$. Then we would have to endow the surface  $r=r_{+}$
with charge and current densities in such a way that the real current outside the horizon
and this fictitious current on the horizon together can complete the circuit. Then the 
task of constructing the horizon surface current can be described as a mathematical
problem as follows : ``Given the bulk current $J^{\mu}(v,r,\theta,\phi)$ such that
$\nabla_{\mu}J^{\mu} = 0$, find a complementary boundary (surface) current $j^{\mu}$
on the surface  $r=r_{+}$ such that $I^{\mu}\equiv [J^{\mu}Y(r-r_{+})+j^{\mu}]$
(where $Y(r)$ is the Heaviside function defined by $dY(r)=\delta(r)dr$) is conserved.''
And in this problem, a crucial point to be noted is that the conservation of the bulk current
$J^{\mu}$ is ensured by the field equation. Obviously then, what changes from the ordinary
Maxwell theory case to the present BI theory case is that now the conservation of $J^{\mu}$
is secured by the inhomogeneous BI field equation instead of the Maxwell equation, i.e.,
\begin{eqnarray}
\nabla_{\nu}G^{\mu\nu} = 4\pi J^{\mu}
\end{eqnarray}
implies $\nabla_{\mu}J^{\mu}=\nabla_{\mu}\nabla_{\nu}G^{\mu\nu}/4\pi=0$ outside the
horizon. Then the condition for the conservation of the {\it total} current $I^{\mu}$
reads
\begin{eqnarray}
0 &=& \nabla_{\mu}I^{\mu} = \nabla_{\mu}[J^{\mu}Y(r-r_{+})+j^{\mu}] \nonumber \\
&=& {1\over 4\pi}(\nabla_{\nu}G^{\mu\nu})\left({x_{\mu}\over r}\right)\delta (r-r_{+})
 + \nabla_{\mu}j^{\mu} 
\end{eqnarray}
where we used $\nabla_{\nu}G^{\mu\nu} = 4\pi J^{\mu}$, $\nabla_{\mu}J^{\mu}=0$ and
$\partial_{\mu}Y=(x_{\mu}/r)\delta (r-r_{+})$. Obviously, this equation is solved by
the complementary surface current given as
\begin{eqnarray}
j^{\mu} &=& {1\over 4\pi}G^{\mu\nu}(\partial_{\nu}r)\delta (r-r_{+}) \nonumber \\
&\equiv& {1\over 4\pi}G^{\mu r}\delta (r-r_{+}).
\end{eqnarray}
Further, it is convenient to introduce a ``Dirac distribution'' $\delta_{H}$ on the
horizon normalized with respect to the time at infinity $v$ and the local proper
area $dA$ such that 
\begin{eqnarray}
\int d^4x \sqrt{g} ~\delta_{H}\delta(v-v_{0})f(v,r,\theta, \phi) =
\int_{H} dA ~f(v_{0},r_{+},\theta, \phi)
\end{eqnarray}
which, then, yields
\begin{eqnarray}
\delta_{H} = {(r^2_{+}+a^2)\over \Sigma_{+}}\delta (r-r_{+})
\end{eqnarray}
where we used $\sqrt{g}=\Sigma \sin \theta$ and $dA=(r^2_{+}+a^2)\sin \theta d\theta
d\phi$. Finally, then, the complementary surface current 4-vector on the horizon can
be written as $j^{\mu}=\kappa^{\mu}\delta_{H}$ with
\begin{eqnarray}
\kappa^{\mu} = {1\over 4\pi}{\Sigma_{+}\over (r^2_{+}+a^2)}G^{\mu r}_{+}.
\end{eqnarray}
As usual, what matters is the identification of ``physical'' (i.e., finite and
non-zero) components of this current 4-vector (i.e., the horizon charge and current
density) as seen by an observer in our quasi-orthonormal tetrad frame.
And they can be computed, using the dual of Damour's mixed tetrad given in eq.(16), in
a straightforward manner as 
\begin{eqnarray}
\sigma &=& \kappa^{0} = \kappa^{\mu}e^{0}_{\mu}|_{r_{+}} = {1\over 4\pi}[G^{vr}_{+}-a\sin^2 \theta 
G^{\phi r}_{+}] \nonumber \\
&=& {1\over 4\pi}\left[{(r^2_{+}+a^2)\over \Sigma_{+}}G_{rv} + {a\over \Sigma_{+}}G_{r\phi}\right]
= {1\over 4\pi}D_{\hat{r}}, \nonumber \\
\kappa^{\hat{r}} &=& \kappa^{1} = \kappa^{\mu}e^{1}_{\mu}|_{r_{+}} = 0, \\
\kappa^{\hat{\theta}} &=& \kappa^{2} = \kappa^{\mu}e^{2}_{\mu}|_{r_{+}} = {1\over 4\pi}
{\Sigma^{3/2}_{+}\over (r^2_{+}+a^2)}G^{\theta r}_{+} \nonumber \\
&=& {1\over 4\pi}\left[{1\over \Sigma^{1/2}_{+}}G_{\theta v} + {a\over \Sigma^{1/2}_{+}(r^2_{+}+a^2)}
G_{\theta \phi}\right] = {1\over 4\pi}D_{\hat{\theta}}, \nonumber \\
\kappa^{\hat{\phi}} &=& \kappa^{3} = \kappa^{\mu}e^{3}_{\mu}|_{r_{+}} = {1\over 4\pi}
{\Sigma^{1/2}_{+}\over (r^2_{+}+a^2)}[-a\sin \theta G^{vr}_{+} + (r^2_{+}+a^2)\sin \theta 
G^{\phi r}_{+}] \nonumber \\
&=& {1\over 4\pi}{\Sigma^{1/2}_{+}\over (r^2_{+}+a^2)\sin \theta}G_{\phi v} = 
{1\over 4\pi}D_{\hat{\phi}} \nonumber
\end{eqnarray}
where the subscript ``$+$'' denotes the value at the horizon $r=r_{+}$ and we compared these
equations with eq.(37) to relate the surface charge and current densities to the components of
electric displacement on the horizon.

\begin{center}
{\rm \bf V. Ohm's law, Gauss' law, and Ampere's law}
\end{center}

We now are in the position to demonstrate that, as results of central significance, a set of three
relations, at the horizon, between the fields ($D_{i}=E_{i}$, $H_{i}=B_{i}$) and the surface charge and
current densities ($\sigma=\kappa^{0}$, $\kappa^{i}$) that can be thought of as Ohm's law, Gauss'
law and Ampere's law valid at the horizon of a rotating Kerr black hole. First, notice that
\begin{eqnarray}
D_{\hat{\theta}} = 4\pi \kappa^{\hat{\theta}}, ~~~D_{\hat{\phi}} = 4\pi \kappa^{\hat{\phi}}.
\end{eqnarray}
These relations can be rewritten in a vector notation in a tangent space to the horizon as
\begin{eqnarray}
\vec{D}_{H} = 4\pi \vec{\kappa} ~~~{\rm or} ~~~\vec{E}_{H} = 4\pi \vec{\kappa}
\end{eqnarray}
and hence can be interpreted as the ``Ohm's law''. Namely, this relation precisely takes on the
form of a non-relativistic Ohm's law for a conductor and hence implies that if we endow the 
horizon with some charge and current densities which are to be determined by the surrounding
external electromagnetic field $F_{\mu\nu}$, then the horizon behaves as if it is a conductor
with finite surface resistivity of 
\begin{eqnarray}
\rho = 4\pi \simeq 377 ({\rm ohms}).
\end{eqnarray}
The derivation of Ohm's law and this value of surface resistivity has been performed first
by Damour [6] and by Znajek [5] independently in the context of standard Maxwell theory. 
Thus what is indeed remarkable here is that the Ohm's law above and the value of horizon's
surface resistivity ($4\pi$) remain unchanged even when we replace the Maxwell theory by the
BI theory of electrodynamics. This result cannot be naturally anticipated but close inspection
reveals that it can be attributed to the peculiar structure of highly non-linear inhomogeneous
BI field equation given in eqs.(30) and (27) which, at the horizon, shows some magic such that
there the ($\vec{D}$, $\vec{H}$) fields become exactly the same as ($\vec{E}$, $\vec{B}$) 
as seen by a local observer in Damour's tetrad frame respectively as can be checked from eqs.(34) 
and (38) (or (40)). As Damour [6] pointed out, this result constitutes a clear confirmation of
Carter's assertion [10] that a black hole is analogous to an ordinary object having finite
viscosity and electrical conductivity. Next, we also notice that 
\begin{eqnarray}
D_{\hat{r}} = 4\pi \sigma ~~~{\rm or} ~~~E_{\hat{r}} = 4\pi \sigma
\end{eqnarray}
which evidently can be identified with the surface version of Gauss' law. It says that the
fictitious surface charge density we assumed on the horizon plays the role of terminating
the normal components of all electric fields that pierce the horizon just as we want it to.
Lastly, if we combine the radiative ingoing boundary condition at the horizon that we 
obtained earlier, $\vec{H}_{H} = \vec{D}_{H} \times \hat{n}$ (or 
$\vec{B}_{H} = \vec{E}_{H} \times \hat{n}$) and the Ohm's law above, $\vec{D}_{H}=4\pi \vec{\kappa}$
(or $\vec{E}_{H} = 4\pi \vec{\kappa}$), we end up with the third relation
\begin{eqnarray}
\vec{H}_{H} = 4\pi (\vec{\kappa} \times \hat{n}) ~~~{\rm or} 
~~~\vec{B}_{H} = 4\pi (\vec{\kappa} \times \hat{n})
\end{eqnarray}
which may be viewed as the surface version of Ampere's law. Again, consistently with our
motivation for introducing fictitious current density on the horizon, this relation
indicates that the current density we assumed plays the role of terminating any tangential
components of all magnetic fields penetrating the horizon. To summarize, for the reason stated
earlier, even the highly non-linear BI theory of electrodynamics leads to the same horizon
boundary conditions eqs.(41), (50), (52), and (53) as those in the standard Maxwell theory 
and indeed this set of four curious boundary conditions on the horizon actually have
provided the motivation for the proposal of membrane paradigm [11] of black holes later on.

\begin{center}
{\rm \bf VI. Joule's law or Ohmic dissipation at the horizon}
\end{center}

Perhaps one of the most intriguing consequences of assuming the existence of fictitious charge
and current densities on the horizon would be that if we choose to do so, the horizon behaves
as if it is a conductor with finite conductivity as we stressed in the previous section.
Since it is the surrounding external electromagnetic field that drives the surface currents
on the horizon, one might naturally wonder what would happen to the Joule heat generated
when those currents work against the surface resistance and how it would be related to the 
electromagnetic energy going down the hole through the horizon. Znajek and Damour also 
provided a simple and natural answer to this question. Namely, they showed in a consistent
and elegant manner that the total electromagnetic energy flux (i.e., the Poynting flux)
into the rotating Kerr hole through the horizon is indeed precisely the same as the amount
of Joule heat produced by the surface currents when they work against the surface resistivity
of $4\pi $. In the following, we shall demonstrate again along the same line of formulation
as Damour that indeed the same is true even in the context of BI theory of electrodynamics.
It is well-known that for a stationary, axisymmetric black hole spacetime with the
horizon-orthogonal Killing field
\begin{eqnarray}
\chi^{\mu} = (\partial /\partial v)^{\mu}+\Omega_{H}(\partial /\partial \phi)^{\mu} \equiv
\xi^{\mu} + \Omega_{H}\psi^{\mu}
\end{eqnarray}
the mass-energy and the angular momentum flux into the hole through the horizon are given
respectively by
\begin{eqnarray}
{dM\over dv} &=& \int_{H}dA ~T^{\mu}_{\nu}\xi^{\nu}\chi_{\mu} = \int_{H}dA ~T^{\mu}_{v}\chi_{\mu}, \\
{dJ_{z}\over dv} &=& -\int_{H}dA ~T^{\mu}_{\nu}\psi^{\nu}\chi_{\mu} = -\int_{H}dA 
~T^{\mu}_{\phi}\chi_{\mu} \nonumber
\end{eqnarray}
where $dA = (r^{2}_{+}+a^2)\sin \theta d\theta d\phi$ is again the area element on the
horizon and $T^{\mu}_{\nu}$ is the matter energy-momentum tensor at the horizon. Now, combining
these with the 1st law of black hole thermodynamics [16]
\begin{eqnarray}
dQ = {1\over 8\pi}\hat{\kappa}_{H}dA = dM - \Omega_{H}dJ_{z}
\end{eqnarray}
where $dQ$ denotes the heat dissipated in the hole {\it not} charge (recall that we only consider
here {\it uncharge} Kerr black hole)
with $\hat{\kappa}_{H}$ being the surface gravity [16] of the hole, one gets
\begin{eqnarray}
{dQ\over dv} &=& {dM\over dv} - \Omega_{H}{dJ_{z}\over dv} \\
&=& \int_{H}dA ~T^{\mu}_{\nu}(\xi^{\nu} + \Omega_{H}\psi^{\nu})\chi_{\mu}
= \int_{H}dA ~T^{\mu}_{\nu}\chi^{\nu}\chi_{\mu}. \nonumber
\end{eqnarray}
Perhaps a word of caution might be relevant here. As we mentioned earlier, we are only
interested in the ``test'' electromagnetic field whose dynamics is governed particularly
by the BI theory in the ``background'' of uncharged Kerr black hole spacetime which is a
solution to the vacuum Einstein equation. Therefore, as long as we confine our concern
to the case with {\it uncharged} Kerr black hole physics, the 1st law of black hole
thermodynamics given above still remains to be valid. If, instead, one is interested in
the case with charged, rotating black hole physics, one would have to deal with the
full, coupled Einstein-BI theory context and then there the associated 1st law should
get modified to the extended version like the one given by Rasheed [17] recently.
Now, since the ``matter'' for the case at hand is the BI electromagnetic field, we have
at the horizon
\begin{eqnarray}
T_{\mu\nu}\chi^{\mu}\chi^{\nu}|_{r_{+}} 
&=& {1\over 4\pi}\left\{\beta^2 (1-R)\chi^{\alpha}\chi_{\alpha} + {1\over R}
\left[F_{\mu\alpha}F_{\nu}^{\alpha}-{1\over 4\beta^2}(F_{\alpha\beta}
\tilde{F}^{\alpha\beta})F_{\mu\alpha}\tilde{F}_{\nu}^{\alpha}\chi^{\mu}\chi^{\nu}
\right]\right\}|_{r_{+}} \nonumber \\
&=& {1\over 4\pi}(F_{\mu\alpha}F_{\nu}^{\alpha})\chi^{\mu}\chi^{\nu}|_{r_{+}}
\end{eqnarray}
where $R$ is as defined earlier and in the second line we used that at the horizon
where $g_{\alpha \beta}\chi^{\alpha}\chi^{\beta}=\chi^{\alpha}\chi_{\alpha}=0$,
\begin{eqnarray}
F_{\alpha \beta}F^{\alpha \beta} &=& -2(E_{\alpha}E^{\alpha}-B_{\alpha}B^{\alpha})
= -2(D_{\alpha}D^{\alpha}-H_{\alpha}H^{\alpha})=0, \nonumber \\
F_{\alpha \beta}\tilde{F}^{\alpha \beta} &=& 4E_{\alpha}B^{\alpha}= 4D_{\alpha}
H^{\alpha}=0, ~~~~{\rm and ~~hence}  \nonumber \\
R &=& \left[1+{1\over 2\beta^2}(F_{\alpha\beta}F^{\alpha\beta})-{1\over 16\beta^4}
(F_{\alpha\beta}\tilde{F}^{\alpha\beta})^2\right]^{1/2}=1 \nonumber
\end{eqnarray}
which also yields, at the horizon, $G_{\mu\nu}=F_{\mu\nu}$.
Recall that in the standard Maxwell theory,
\begin{eqnarray}
T_{\mu\nu} = {1\over 4\pi}[F_{\mu\alpha}F_{\nu}^{\alpha}-{1\over 4}g_{\mu\nu}
(F_{\alpha \beta}F^{\alpha \beta})]
\end{eqnarray}
and thus at the horizon, $T_{\mu\nu}\chi^{\mu}\chi^{\nu}|_{r_{+}}=(1/4\pi)
(F_{\mu\alpha}F_{\nu}^{\alpha})\chi^{\mu}\chi^{\nu}|_{r_{+}}$, which is the same
as its counterpart in BI theory obtained above. This means that, at the horizon,
the amount of total electromagnetic energy flux into the hole turns out to be the 
same and hence indistinguishable between Maxwell theory and BI theory. Further,
\begin{eqnarray}
T_{\mu\nu}\chi^{\mu}\chi^{\nu}|_{r_{+}} &=& {1\over 4\pi}
(F_{\mu\alpha}F_{\nu}^{\alpha})\chi^{\mu}\chi^{\nu}|_{r_{+}} \\
&=& {1\over 4\pi}\left\{\left[{\Sigma^{1/2}_{+}\over (r^2_{+}+a^2)\sin \theta}F_{\phi v}
\right]^2 + \left[{1\over \Sigma^{1/2}_{+}}G_{\theta v} + {a\over \Sigma^{1/2}_{+}(r^2_{+}+a^2)}
G_{\theta \phi}\right]^2\right\} \nonumber \\
&=& 4\pi [(\kappa^{\hat{\phi}})^2 + (\kappa^{\hat{\theta}})^2] =  4\pi (\vec{\kappa})^2 \nonumber  
\end{eqnarray}
where we used $G_{\mu\nu}=F_{\mu\nu}$ and $\kappa^{\hat{r}}=0$ at the horizon. Thus, finally
we end up with
\begin{eqnarray}
{dQ\over dv} &=& \int_{H}dA ~T_{\mu \nu}\chi^{\mu}\chi^{\nu} \\
&=& 4\pi \int_{H}dA ~(\vec{\kappa})^2 = \int_{H}dA ~(\vec{E}_{H}\cdot \vec{\kappa}) \nonumber  
\end{eqnarray}
where we used the Ohm's law $\vec{E}_{H}= 4\pi \vec{\kappa}$, we obtained earlier. As we
promised to demonstrate, clearly this is the Joule's law which is again precisely the same as its
Maxwell theory counterpart originally obtained by Znajek [5] and by Damour [6] and implies that 
the absorption of electromagnetic energy by Kerr holes through the horizon can be translated into
an equivalent picture in which the holes gain energy by absorbing Joule heat (or Ohmic dissipation)
generated when the surface current $\vec{\kappa}$ driven by the electric field $\vec{E}_{H}$ works
against the surface resistivity of $4\pi $. And as before, what is remarkable is the fact that
even if we replace the Maxwell theory by the highly non-linear BI electrodynamics, the physics
of the horizon such as this horizon thermodynamics as well as the horizon boundary conditions
remain unchanged. And as we pointed out earlier, this has much to do with the nature of Damour's
quasi-orthonormal tetrad frame (i.e., its half-null structure) in ingoing Kerr coordinates.

\begin{center}
{\rm \bf VII. Concluding Remarks}
\end{center}

In the present work, we have explored the interaction of electromagnetic fields with a rotating 
(Kerr) black hole in the context of Born-Infeld (BI) theory of electromagnetism
and particularly we have derived BI theory versions of the four horizon boundary
conditions of Znajek and Damour. Interestingly enough,
as far as we employ the same local null tetrad frame as the one adopted in the
works by Damour and by Znajek, we ended up with exactly the same four  
horizon boundary conditions despite the shift of the electrodynamics theory
from a linear Maxwell one to a highly non-linear BI one. As we have seen in
the text, this curious and unexpected result could be attributed to the fact that
the concrete structure of BI equations happens to be 
such that it is indistinguishable {\it at the horizon} to a local observer, say, in Damour's
local tetrad frame from that of standard Maxwell theory.
Finally, we have a word of caution to avoid a possible confusion the potential readers might have.
Namely, again we point out that in all the calculations involved in this work to read off physical 
components of tensors such as Maxwell field tensor and current 4-vector, we strictly used
the quasi-orthonormal tetrad given in eqs.(15) and (16) and nothing else. In this sense,
our choice of local tetrad frame was slightly different from that in the original work of
Damour [6] in which he introduced, particularly on the 2-dimensional, $v=const.$ section of
the event horizon, some other orthonormal basis (slightly different from $\{e^{\mu}_{2},
~e^{\mu}_{3}\}$ given in eq.(15)) specially adapted to the ``intrinsic geometry'' of the 
$v=const.$ section of the horizon and used them to project out physical components of tensors.
As such, any deviation of the results one may find in the expressions for the electric field,
magnetic field and surface charge and current densities appeared in the text of the present work
from their counterparts in the original work of Damour can be attributed to this slightly
different choices of the local tetrad vectors. This discrepancy, however, is insensitive to
the physical nature of this study of the horizon boundary conditions that we try to deliver 
in the present work.

\vspace*{1cm}

\begin{center}
{\rm\bf Acknowledgements}
\end{center}

This work was supported in part by the Brain Korea 21 Project. HKL and CHL are also supported
in part by the interdisciplinary research program of the KOSEF, Grant No. 1999-2-003-5.

\vspace*{2cm}

\begin{center}
{\rm\bf Appendix : An Introduction to Born-Infeld Electrodynamics}
\end{center}

The Born-Infeld (BI) theory may be thought of as a highly nonlinear generalization of or
a non-trivial alternative to the standard Maxwell theory of electromagnetism. 
Here, we would like to present an introductory review of BI theory of electrodynamics particularly
in modern field theory perspective. As usual, we begin with the action for this BI theory
which is given, in 4-dimensions, by (in $MKS$ unit) 
\begin{eqnarray}
S &=& \int d^4x \left\{ \beta^2\left[1 - \sqrt{-det(\eta_{\mu\nu} + 
{1\over \beta}F_{\mu\nu})}\right]
+ j^{\mu}A_{\mu}\right\} \\
&=& \int d^4x \left\{ \beta^2\left[1 - \sqrt{1 + {1\over 2\beta^2}F_{\mu\nu}F^{\mu\nu} -
{1\over 16\beta^4}(F_{\mu\nu}\tilde{F}^{\mu\nu})^2}\right] + j^{\mu}A_{\mu}\right\} \nonumber
\end{eqnarray}
where ``$\beta$'' is a generic parameter of the theory having the dimension $dim[\beta] =
dim[F_{\mu\nu}] = +2$. It probes the degree of deviation of BI gauge theory from the standard 
Maxwell theory and obviously $\beta \rightarrow \infty$ limit corresponds to the standard
Maxwell theory. Again, extremzing this action with respect to $A_{\mu}$ yields the dynamical
BI field equation 
\begin{eqnarray}
\partial_{\mu}\left[{{F^{\mu\nu} - {1\over 4\beta^2}(F_{\alpha\beta}\tilde{F}^{\alpha\beta})
\tilde{F}^{\mu\nu}} \over {\sqrt{1 + {1\over 2\beta^2}F_{\alpha\beta}F^{\alpha\beta} - 
{1\over 16\beta^4}(F_{\alpha\beta}\tilde{F}^{\alpha\beta})^2}}}\right] = - j^{\nu}.
\end{eqnarray}
In addition to this, there is a supplementary equation coming from an identity satisfied by
the abelian gauge field strength tensor,
$\partial_{\lambda}F_{\mu\nu} + \partial_{\mu}F_{\nu\lambda} + \partial_{\nu}F_{\lambda\mu}=0.$  
This is the Bianchi identity which is just a geometrical equation and in terms of the Hodge
dual field strength, $\tilde{F}^{\mu\nu} = {1\over 2}\epsilon^{\mu\nu\alpha\beta}F_{\alpha\beta}$,
it can be written as
\begin{eqnarray}
\partial_{\mu}\tilde{F}^{\mu\nu} = 0
\end{eqnarray}
And it seems noteworthy that the field equation for $A_{\mu}$ in eq.(63) is the dynamical field
equation which gets determined by the concrete nature of the gauge theory action such as the
one in eq.(62). The Bianchi identity in eq.(64), on the other hand, is simply a geometrical
identity and is completely independent of the choice of the context of the gauge theory.
Further, if one wishes to decompose these covariant equations, use $\partial^{\mu} = 
(-\partial/\partial t, \nabla_{i})$, 
$\partial_{\mu}=\eta_{\mu\nu}\partial^{\nu} = (\partial/\partial t, \nabla_{i})$
(namely, we use the sign convention, $\eta_{\mu\nu} = diag(-1, 1, 1, 1)$), $A^{\mu}=(\phi, A^{i})$
and the field identification, $E_{i}=F_{i0}$, $B_{i}={1\over 2}\epsilon_{ijk}F^{jk}$ or
$F_{ij}=\epsilon_{ijk}B^{k}$, and write them in terms of $\vec{E}$ and
$\vec{B}$ fields, to get
\begin{eqnarray}
&\nabla & \cdot \left[{1\over R}(\vec{E} + {1\over \beta^2}(\vec{E}\cdot \vec{B})\vec{B})\right]
= \rho_{e}, \\
&\nabla & \times \left[{1\over R}(\vec{B} - {1\over \beta^2}(\vec{E}\cdot \vec{B})\vec{E})\right] 
- {\partial \over \partial t}
\left[{1\over R}(\vec{E} + {1\over \beta^2}(\vec{E}\cdot \vec{B})\vec{B})\right] = \vec{j}_{e}
\nonumber
\end{eqnarray}
where $R \equiv \sqrt{1 + {1\over 2\beta^2}F_{\alpha\beta}F^{\alpha\beta} -
{1\over 16\beta^4}(F_{\alpha\beta}\tilde{F}^{\alpha\beta})^2} = \sqrt{1 - {1\over \beta^2}
(\vec{E}^2 - \vec{B}^2) - {1\over \beta^4}(\vec{E}\cdot \vec{B})^2}$ for the dynamical BI field
equation and 
\begin{eqnarray}
\nabla \cdot \vec{B} = 0,  ~~~\nabla \times \vec{E} + {\partial \vec{B}\over \partial t} = 0
\end{eqnarray}
for the geometrical Bianchi identity and where we used $F_{\mu\nu}F^{\mu\nu} = -2(\vec{E}^2-\vec{B}^2)$
and $F_{\mu\nu}\tilde{F}^{\mu\nu} = 4\vec{E}\cdot \vec{B}$. 
We now start with some electrostatics described by this BI theory. As a simplest exercise, we look
for the solution to these BI equations that represents a static electric monopole field.
Next, the static electric monopole. It can be obtained from one of the dynamical field equations
$\nabla \cdot [\left\{\vec{E}+(\vec{E}\cdot \vec{B})\vec{B}/\beta^2\right\}/R] = e\delta^3(\vec{r})$
with $R \equiv \left\{1 - (\vec{E}^2-\vec{B}^2)/\beta^2 - (\vec{E}\cdot \vec{B})^2/\beta^4\right\}^{1/2}$.
Again for $\vec{r}\neq 0$, and in spherical-polar coordinates, this equation becomes
$[\partial_{r}(r^2 \sin\theta \hat{E}_{r}) + \partial_{\theta}(r \sin\theta \hat{E}_{\theta}) +
\partial_{\phi}(r \hat{E}_{\phi})]/r^2\sin\theta = 0$ with
$\hat{E}_{i} \equiv [E_{i}+(\vec{E}\cdot \vec{B})B_{i}/\beta^2]/R$. In the absence of the
magnetic field, $\hat{E}_{i} = E_{i}/\sqrt{1-\vec{E}^2/\beta^2}$ and then the above equation is
solved by [15]
\begin{eqnarray}
E_{r} = {e\over {4\pi \sqrt{r^4 + \left({e\over 4\pi \beta}\right)^2}}},
~~~E_{\theta} = E_{\phi} = 0.
\end{eqnarray}
Since the static electric monopole field is not singular
as $r \rightarrow 0$, the energy stored in the field of electric point charge could be finite and this
point seems to be consistent with the consideration of finiteness of energy, which is one of the
motivations to propose this BI electrodynamics when it was first devised. 
Thus to see if this is indeed the case, consider the energy-momentum tensor of this BI theory
\begin{eqnarray}
T_{\mu\nu} = \beta^2 (1 - R)\eta_{\mu\nu} + {1\over R}\left[F_{\mu\alpha}F_{\nu}^{\alpha} -
{1\over 4\beta^2}(F_{\alpha\beta}F^{\alpha\beta})F_{\mu\alpha}\tilde{F}_{\nu}^{\alpha}\right]
\end{eqnarray}
with $R$ as given earlier. The energy density stored in the electromagnetic field can then be read off
as
\begin{eqnarray}
T_{00} = \beta^2 \left[{1 + {1\over \beta^2}\vec{B}^2 \over {\sqrt{1 - {1\over \beta^2}
(\vec{E}^2 - \vec{B}^2) - {1\over \beta^4}(\vec{E}\cdot \vec{B})^2}}} - 1 \right]
\end{eqnarray}
which does reduce to its Maxwell theory's counterpart $(\vec{E}^2 + \vec{B}^2)/2$ in the limit
$\beta \rightarrow \infty$ as it should. We are ready to calculate the energy density stored in the
electric field generated by the electric charge $e$.
Using $\vec{E} = \left\{e/4\pi \sqrt{r^4 + (e/4\pi\beta)^2}\right\}\hat{r}$, one gets
\begin{eqnarray}
T^{E}_{00} = \beta^2 \left[{1\over \sqrt{1 - {1\over \beta^2}\vec{E}^2}} - 1\right] =
\beta^2 \left[\sqrt{1 + {e^2\over (4\pi\beta)^2}{1\over r^4}} - 1\right].
\end{eqnarray}
Then the electric monopole energy can be evaluated in a concrete manner as [15] 
\begin{eqnarray}
E &=& \int d^3x T^{E}_{00} = \int^{\infty}_{0}dr \beta^2 \left[\sqrt{(4\pi r^2)^2 + {e^2\over \beta^2}}
- 4\pi r^2\right] \nonumber \\
&=& \sqrt{{\beta e^3\over 4\pi}}\int^{\infty}_{0}dy \left[\sqrt{y^4 + 1} - y^2\right] \\
&=& \sqrt{{\beta e^3\over 4\pi}}{\pi^{3/2}\over 3\Gamma({3\over 4})^2} =
1.23604978 \sqrt{{\beta e^3\over 4\pi}} \nonumber 
\end{eqnarray}
where $y^2 = (4\pi \beta/e^2)r^2$ and in the $y$-integral, integration by part and the elliptic
integral have been used. Remarkably, this energy is indeed {\it finite} as Born and Infeld hoped
when they constructed this theory and if one takes the Maxwell theory limit $\beta \rightarrow
\infty$, one recovers divergent energy for a point electric charge as expected.
Lastly we turn to some electrodynamics governed by this BI theory.
In the dynamical BI field equations given earlier in eq.(69), we define, for the sake of convenience 
of the formulation, the ``electric displacement'' $\vec{D}$ and the ``magnetic field'' 
$\vec{H}$ in terms of the fundamental fields $\vec{E}$ and $\vec{B}$ as
\begin{eqnarray}
\vec{D} = {1\over R}\left\{\vec{E} + {1\over \beta^2}(\vec{E}\cdot \vec{B})\vec{B}\right\},
~~~\vec{H} = {1\over R}\left\{\vec{B} - {1\over \beta^2}(\vec{E}\cdot \vec{B})\vec{E}\right\}  
\nonumber
\end{eqnarray}
where $R = \sqrt{1 - {1\over \beta^2}(\vec{E}^2 - \vec{B}^2) - {1\over \beta^4}(\vec{E}\cdot \vec{B})^2}$
is as defined earlier. Then the BI equations take the form 
\begin{eqnarray}
\nabla \cdot \vec{D} &=& \rho_{e},  ~~~\nabla \times \vec{H} - {\partial \vec{D}\over \partial t}
= \vec{j}_{e} \\
\nabla \cdot \vec{B} &=& 0,  ~~~\nabla \times \vec{E} + {\partial \vec{B}\over \partial t}
= 0. \nonumber
\end{eqnarray}
Now $\vec{E}\cdot ({\rm Ampere's ~law ~eq.}) - \vec{H}\cdot ({\rm Faraday's ~induction ~law ~eq.})$ yields
\begin{eqnarray}
\vec{H}\cdot (\nabla \times \vec{E}) - \vec{E}\cdot (\nabla \times \vec{H}) =
- \vec{H}\cdot {\partial \vec{B} \over \partial t} - \vec{E}\cdot {\partial \vec{D} \over \partial t}
- \vec{j}_{e}\cdot \vec{E}. \nonumber
\end{eqnarray}
Further using
\begin{eqnarray}
\vec{H}\cdot (\nabla \times \vec{E}) - \vec{E}\cdot (\nabla \times \vec{H}) = \nabla \cdot 
(\vec{E}\times \vec{H}), 
~~~- \vec{H}\cdot {\partial \vec{B} \over \partial t} - \vec{E}\cdot {\partial \vec{D} \over \partial t}
= - {\partial \over \partial t}T_{00} \nonumber
\end{eqnarray}
where $T_{00}$ is the energy density stored in the electromagnetic field in BI theory given in eq.(69),
one arrives at the familiar local energy conservation equation
\begin{eqnarray}
\nabla \cdot \vec{S} + {\partial u \over \partial t} = - \vec{j}_{e}\cdot \vec{E}
\end{eqnarray}
where $u = T_{00}$ is the energy density, $\vec{S} = \vec{E} \times \vec{H}$ is the ``Poynting vector''
representing the local energy flow per unit time per unit area and 
$- \vec{j}_{e}\cdot \vec{E}$ on the right hand side is the power dissipation
per unit volume. In particular for $\vec{j}_{e}\cdot \vec{E} = 0$, one gets
\begin{eqnarray}
\nabla \cdot \vec{S} + {\partial u \over \partial t} = 0 \nonumber
\end{eqnarray}
which is the equation of continuity for electromagnetic energy density with
the BI theory version of the Poynting vector given by [15]
\begin{eqnarray}
\vec{S} = \vec{E} \times \vec{H} = {\vec{E} \times \vec{B} \over
\sqrt{1 - {1\over \beta^2}(\vec{E}^2 - \vec{B}^2) - {1\over \beta^4}(\vec{E}\cdot \vec{B})^2}}
\end{eqnarray}
which obviously reduces to its Maxwell theory counterpart $\vec{S} = \vec{E} \times \vec{B}$ in
the limit $\beta \rightarrow \infty$.

\noindent

\begin{center}
{\rm\bf References}
\end{center}

\begin{description}

\item {[1]} E. E. Salpeter, Astrophys. J. {\bf 140}, 796 (1964) ;
            Ya. B. Zeldovich, Soviet Physics - Doklady, {\bf 9}, 195 (1964) ;
            JETP Lett. {\bf 14}, 180 (1971).
\item {[2]} R. Penrose, Nuovo. Cim. {\bf 1}, 252 (1969) ;
            W. H. Press and S. A. Teukolsky, Nature {\bf 238}, 211 (1972).
\item {[3]} R. Ruffini and J. R. Wilson, Phys. Rev. {\bf D12}, 2959 (1975) ;
            T. Damour, Ann. N.Y. Acad. Sci. {\bf 262}, 113 (1975).
\item {[4]} R. D. Blandford and R. L. Znajek, Mon. Not. R. Astron. Soc. {\bf 179}, 433 (1977) ;
            as nice review articles, also see, D. Macdonald and K. S. Thorne, {\it ibid.} {\bf 198},
            345 (1982) ; H. K. Lee, R. A. M. Wijers, and G. E. Brown, Phys. Rep. {\bf 325}, 83
            (2000).
\item {[5]} R. L. Znajek,  Mon. Not. R. Astron. Soc. {\bf 185}, 833 (1978).
\item {[6]} T. Damour, Phys. Rev. {\bf D18}, 3598 (1978).
\item {[7]} R. S. Hanni and R. Ruffini, Phys. Rev. {\bf D8}, 3259 (1973).
\item {[8]} C. Kouveliotou, Astrophys. J. {\bf 510}, L115 (1999).
\item {[9]} M. Born, Proc. R. Soc. London {\bf A143}, 410 (1934) ; 
            M. Born and M. Infeld, {\it ibid.} {\bf A144}, 425 (1934) ;
            P. A. M. Dirac,  {\it ibid.} {\bf A268}, 57 (1962).
\item {[10]} B. Carter, in {\it Black Holes}, edited by B. DeWitt and C. DeWitt (Gordon and Breach,
            New York, 1973).
\item {[11]} K. S. Thorne, R. H. Price, and D. A. Macdonald, {\it Black Holes: The Membrane
             Paradigm}, (Yale University Press, New Haven and London, 1986).
\item {[12]} J. B. Hartle and S. W. Hawking, Commun. Math. Phys. {\bf 27}, 283 (1972).
\item {[13]} S. A. Teukolsky,  Astrophys. J. {\bf 185}, 635 (1973).
\item {[14]} W. Kinnersley, J. Math. Phys. {\bf 10}, 1195 (1969).
\item {[15]} H. Kim,  Phys. Rev. {\bf D61}, 085014 (2000) and references therein.
\item {[16]} See, for example, R. M. Wald, {\it Quantum Field Theory in Curved Spacetime and 
        Black Hole Thermodynamics}, (The University of Chicago Press, Chicago and London, 1994).
\item {[17]} D. A. Rasheed, DAMTP Preprint R97/08, {\it hep-th/9702087}.

\end{description}

\end{document}